\documentclass{elsart}

\journal{Annals of Physics}

\usepackage{graphicx,amsmath}

\newcommand{\intx}{\ensuremath{\int d^4 x}}

\newcommand{\sgn}{\ensuremath{\mathrm{sgn}}}
\newcommand{\intq}{\ensuremath{\int\frac{d^4 q}{(2\pi)^4}}}

\newcommand{\qperp}{\ensuremath{q_\perp}}
\newcommand{\qpara}{\ensuremath{q_\parallel}}
\newcommand{\ppara}{\ensuremath{p_\parallel}}
\newcommand{\gammaperp}{\ensuremath{\gamma_\perp}}
\newcommand{\gammapara}{\ensuremath{\gamma_\parallel}}
\newcommand{\nn}{\nonumber}
\begin{document}
\begin{frontmatter}
\title{Gauge independence and chiral symmetry breaking in a strong magnetic
field}
\author[udel]{C. N. Leung}
\ead{leung@physics.udel.edu} and
\author[udel,tku]{S.-Y. Wang\corauthref{cor}}
\ead{sywang@mail.tku.edu.tw} \corauth[cor]{Corresponding author.}
\address[udel]{Department of Physics and Astronomy, University of
Delaware, Newark, Delaware 19716, USA}
\address[tku]{Department of Physics, Tamkang University, Tamsui, Taipei
25137, Taiwan}

%\date{\today}

\begin{abstract}
The gauge independence of the dynamical fermion mass generated
through chiral symmetry breaking in QED in a strong, constant
external magnetic field is critically examined. We present a
(first, to the best of our knowledge) consistent truncation of the
Schwinger-Dyson equations in the lowest Landau level
approximation. We demonstrate that the dynamical fermion mass,
obtained as the solution of the truncated Schwinger-Dyson
equations evaluated on the fermion mass shell, is manifestly gauge
independent.
\end{abstract}

%\pacs{
%11.30.Rd, %Chiral symmetries
%11.30.Qc, %Spontaneous and radiative symmetry breaking
%12.20.-m %Quantum electrodynamics
%}

%\keywords{Chiral symmetry breaking, Gauge independence, Quantum
%electrodynamics}

\end{frontmatter}

Chiral symmetry breaking in an external magnetic field has attracted
a lot of attention in the past decade. Being inherently a
nonperturbative phenomenon, the generation of a dynamical fermion
mass is usually studied with the help of the Schwinger-Dyson (SD)
equations truncated in certain schemes. The dynamical fermion mass
has been calculated in the literature in several truncation schemes,
such as the
rainbow~\cite{Gusynin:1995gt,Leung:1996qy,Gusynin:1995nb} and the
improved
rainbow~\cite{Gusynin:1995nb,Gusynin:1998zq,Kuznetsov:2002zq}
approximations with a momentum independent fermion self-energy, as
well as their extensions to a momentum dependent fermion
self-energy~\cite{Gusynin:1998zq,Kuznetsov:2002zq,Alexandre:2000nz}.
However, to the best of our knowledge, issues regarding the
consistency of truncation schemes as well as the gauge independence
of the dynamical fermion mass have not been properly addressed in
the previous literature in this field.

The demonstration of gauge independence of physical quantities is of
fundamental importance in gauge
theories~\cite{Nielsen:1975fs,Kobes:1990dc}. In particular, the
gauge independence of physical quantities obtained in a
nonperturbative calculation (e.g., the SD equations) is a highly
nontrivial problem. This is because an infinite subset of diagrams
arising from every order in the loop expansion has to be resummed
consistently, thus leading to potential gauge dependence of physical
quantities whenever not all relevant diagrams are accounted for.
Therefore, we emphasize that in gauge theories no truncation schemes
of the SD equations should be considered consistent unless the gauge
independence of physical quantities calculated therein is
unequivocally demonstrated.

In this article, we critically study dynamical chiral symmetry
breaking in weak coupled QED in a strong, constant external magnetic field.
We demonstrate that there exists a consistent
truncation of the SD equations in the lowest Landau level
approximation. We show that within this consistent truncation scheme
the dynamical fermion mass, obtained as the solution of the
truncated SD equations evaluated \emph{on the fermion mass shell},
is manifestly gauge independent.

We take the constant external magnetic field of strength $H$ in
the $x_3$-direction. The corresponding vector potential is given
by $A_\mu^\mathrm{ext}=(0,0,Hx_1,0)$, where $\mu=0,1,2,3$. In our
convention, the metric has the signature
$g_{\mu\nu}=\mathrm{diag}(-1,1,1,1)$. The SD equations in QED in
an external magnetic field are well-known in the literature. The
equations for the full fermion propagator $G(x,y)$ are given by
\begin{eqnarray}
G^{-1}(x,y)&=&S^{-1}(x,y)+\Sigma(x,y),\label{SDfermion}\\
\Sigma(x,y)&=&ie^2\intx' d^4y' \,\gamma^\mu\,G(x,x')\,
%\nn\\&&\times
\Gamma^\nu(x',y,y')\,\mathcal{D}_{\mu\nu}(x,y'),\label{SDSigma}
\end{eqnarray}
where $S(x,y)$ is the bare fermion propagator in the external
field $A_\mu^\mathrm{ext}$, $\Sigma(x,y)$ is the fermion
self-energy and $\Gamma^\nu(x,y,z)$ is the full vertex. The full
photon propagator $\mathcal{D}_{\mu\nu}(x,y)$ satisfies the
equations
\begin{eqnarray}
\mathcal{D}^{-1}_{\mu\nu}(x,y)&=&D^{-1}_{\mu\nu}(x,y)+\Pi_{\mu\nu}(x,y),\\
\Pi_{\mu\nu}(x,y)&=&-ie^2\,\mathrm{tr}\intx' d^4y'
\,\gamma_\mu\,G(x,x')\,
%\nn\\&&\times
\Gamma_\nu(x',y',y)\,G(y',x),\label{SDphoton}
\end{eqnarray}
where $D_{\mu\nu}(x,y)$ is the free photon propagator
(defined in covariant gauges) and
$\Pi_{\mu\nu}(x,y)$ is the vacuum polarization.

Since the dynamics of fermion pairing in a strong magnetic field
is dominated by the lowest Landau level
(LLL)~\cite{Gusynin:1995gt,Leung:1996qy,Gusynin:1995nb,Gusynin:1998zq},
we will consider the propagation of, as well as radiative
corrections originating only from, fermions occupying the LLL.
This is referred to as the lowest Landau level approximation
(LLLA) in the literature. Consequently, for the purpose of this
article, the fermion propagator and the fermion self-energy in the
SD equations \eqref{SDfermion}-\eqref{SDphoton} will be taken to
be those for the LLL fermion. For notational simplicity, no
separate notation will be introduced.

It is well-known that the SD equations
\eqref{SDfermion}-\eqref{SDphoton} do not form a closed system of
integral equations unless a truncation scheme is employed by
specifying the full vertex $\Gamma^\mu$ in terms of other entities
already appeared in the SD equations.
To this end we will work in
the bare vertex approximation (BVA), in which the vertex corrections
are completely ignored. This is achieved by replacing the full vertex
in the SD equations \eqref{SDfermion}-\eqref{SDphoton} by the bare
one, viz,
\begin{equation}
\Gamma^\mu(x,y,z)=\gamma^\mu\,\delta^{(4)}(x-z)\,\delta^{(4)}(y-z).
\label{BVA}
\end{equation}
This truncation is also known as the (improved) rainbow
approximation and has been employed extensively in the
literature~\cite{Gusynin:1995gt,Leung:1996qy,Gusynin:1995nb,Gusynin:1998zq,Kuznetsov:2002zq,Alexandre:2000nz}.
However, we emphasize that unlike what has usually been done in the
literature, here we will \emph{not} confine ourselves to a
particular gauge (usually the Feynman
gauge)~\cite{Gusynin:1995gt,Leung:1996qy,Gusynin:1995nb}, \emph{nor}
will we make the assumption that the BVA \eqref{BVA} is valid only
in a certain
gauge~\cite{Gusynin:1998zq,Kuznetsov:2002zq,Alexandre:2000nz}.
Instead, it is our aim to prove that the BVA \eqref{BVA} is a
consistent truncation of the SD equations
\eqref{SDfermion}-\eqref{SDphoton} in the LLLA. The dynamical
fermion mass, obtained as the solution of the truncated SD equations
evaluated on the fermion mass shell, is manifestly gauge
independent. In the weak coupling regime that we consider, such a
gauge independent approach allows one to resum consistently an
infinite subset of diagrams that arises from every order in the loop
expansion and whose contributions are of leading order in the gauge
coupling, thus leading to a consistent and reliable calculation of
the dynamical fermion mass.

The main ingredient in the proof of the gauge independence of
physical quantities is the Ward-Takahashi (WT) identity
satisfied by the vertex and the inverse fermion propagator. The WT
identity for the bare vertex \eqref{BVA} takes the form
\begin{equation}
\delta^{(4)}(x-y)\,e^{-iq\cdot x}\,\gamma\cdot q=(e^{-iq\cdot
x}-e^{-iq\cdot y})\,G^{-1}(x,y),\label{WT}
\end{equation}
where $q^\mu$ is the momentum carried by the photon. It was shown in
Ref.~\cite{Ferrer:1998vw} that in order to satisfy the WT identity
in the BVA \eqref{WT}, the LLL fermion self-energy in momentum space
has to be a momentum independent constant. We note that due to an
oversight in Refs.~\cite{Leung:1996qy,Ferrer:1998vw} regarding the
matrix structure in the orthonormal condition of the Ritus $E_p$
functions~\cite{Ritus} for the LLL fermions, the calculations
therein require further investigations. It can be
shown~\cite{Leung:2005yq} that with the correct orthonormal
condition the conclusion obtained in Ref.~\cite{Ferrer:1998vw} on
the WT identity in the BVA \eqref{WT} remains valid within the LLLA.
The reliability of such a momentum independent approximation and,
consequently, of the WT identity in the BVA \eqref{WT} has been
verified in certain gauges in the momentum region relevant to the
dynamics of fermion pairing in a strong magnetic
field~\cite{Gusynin:1998zq,Kuznetsov:2002zq,Alexandre:2000nz}.

As per the WT identity in the BVA \eqref{WT}, we can write the
self-energy for the LLL fermion as $\Sigma(\ppara)=m(\xi)$, where
$\ppara$ is the momentum of the LLL fermion and $m(\xi)$ is a
momentum independent \emph{but} gauge dependent constant, with
$\xi$ being the gauge parameter in covariant gauges. Here and
henceforth, the subscript $\parallel$ ($\perp$) refers to the
longitudinal: $\mu=0,3$ (transverse: $\mu=1,2$) components. It is
noted that $m(\xi)$ depends implicitly on $\xi$ through the full
photon propagator $\mathcal{D}_{\mu\nu}$ in \eqref{SDSigma}. We
emphasize that because of its $\xi$-dependence, $m(\xi)$ should
\emph{not} be taken for granted to be the dynamical fermion mass,
which is a gauge independent physical quantity.

We now begin the proof that the BVA is a consistent truncation of
the SD equations \eqref{SDfermion}-\eqref{SDphoton}, in which
$m(\xi)$ is $\xi$-independent and hence can be identified
unambiguously as the dynamical fermion mass, \emph{if and only if}
the truncated SD equation for the fermion self-energy is evaluated
on the fermion mass shell.

We first recall that, as proved in Ref.~\cite{Kobes:1990dc}, in
gauge theories the singularity structures (i.e., the positions of
poles and branch singularities) of gauge boson and fermion
propagators are gauge independent when all contributions of a
given order of a systematic expansion scheme are accounted for.
Consequently, this means the dynamical fermion mass has to be
determined by the pole of the full fermion propagator obtained in
a consistent truncation scheme.

The full propagator for the LLL fermion is given by
\begin{equation}
G(\ppara)=\frac{1}{\gammapara\cdot\ppara+\Sigma(\ppara)}\,
\Delta[\sgn(eH)],\label{G}
\end{equation}
where $\Delta[\sgn(eH)]=[1+i\gamma^1\gamma^2\,\sgn(eH)]/2$ is the
projection operator on the fermion states with the spin parallel to
the external magnetic field. Assume for the moment that the BVA is a
consistent truncation of the SD equations in the LLLA, such that the
position of the pole of $G(\ppara)$ in \eqref{G} is gauge
independent. In accordance with the WT identity in the BVA
\eqref{WT}, we have
\begin{equation}
\Sigma(\ppara)=\Sigma(\ppara^2=-m^2)=m,\label{Sigma}
\end{equation}
where $m$ is the \emph{gauge independent, physical dynamical
fermion mass}, yet to be determined by solving the truncated SD
equations self-consistently. What remains to be verified in our
proof is the following statements: (i) the truncated vacuum
polarization is transverse; (ii) the truncated fermion self-energy
is gauge independent when evaluated on the fermion mass shell,
$\ppara^2=-m^2$.

We highlight that the fermion mass shell condition is one of the
most important points that has gone unnoticed in the literature,
where the truncated fermion self-energy used to be evaluated off the
fermion mass shell at, say,
$\ppara^2=0$~\cite{Gusynin:1995gt,Leung:1996qy,Gusynin:1995nb,Gusynin:1998zq,Kuznetsov:2002zq,Alexandre:2000nz}.

In terms of \eqref{G} and \eqref{Sigma}, the vacuum polarization
$\Pi_{\mu\nu}(q)$ in the BVA is found to be given
by~\cite{Leung:2005yq}
\begin{eqnarray}
\Pi^{\mu\nu}(q)&=&-\frac{ie^2}{2\pi}\,|eH|\,
\exp\left(-\frac{q_\perp^2}{2|eH|}\right)
\mathrm{tr}\int\frac{d^2p_\parallel}{(2\pi)^2}
%\nn\\&&\times
\gamma^\mu_\parallel\,\frac{1}{\gammapara\cdot\ppara+m}\,
\gamma^\nu_\parallel\nn\\
&&\times\frac{1}{\gammapara\cdot(p-q)_\parallel+m}\Delta[\sgn(eH)].\label{Pi}
\end{eqnarray}
In obtaining \eqref{Pi}, we have made use of the following
properties
\begin{eqnarray}
\Delta[\sgn(eH)]\,\gammapara^\mu\,\Delta[\sgn(eH)]&=&\gammapara^\mu\,\Delta[\sgn(eH)],
\label{Delta1}\\
\Delta[\sgn(eH)]\,\gammaperp^\mu\,\Delta[\sgn(eH)]&=&0.\label{Delta2}
\end{eqnarray}
The presence of $\Delta[\sgn(eH)]$ in \eqref{Pi} is a consequence
of the LLLA, which, as explicitly displayed in \eqref{Pi}, leads
to an effective dimensional
reduction~\cite{Gusynin:1995gt,Gusynin:1995nb}.

With the LLL fermion self-energy given by \eqref{Sigma}, the WT
identity in the BVA \eqref{WT} reduces in momentum space
to~\cite{Leung:2005yq}
\begin{equation}
\gammapara\cdot \qpara=
(\gammapara\cdot\ppara+m)-[\gammapara\cdot(p-q)_\parallel+m],\label{WT2}
\end{equation}
where, due to \eqref{Delta2}, the transverse components
$\gammaperp\cdot\qperp$ on the left-hand side decouple in the LLLA.

Upon using the WT identity in the BVA \eqref{WT2}, one can verify
that $\Pi_{\mu\nu}(q)$ is transverse, i.e., $q^\mu
\Pi_{\mu\nu}(q)=0$. Explicit calculation in dimensional
regularization shows that the $1/\epsilon$ pole corresponding to
an ultraviolet logarithmic divergence cancels, leading to
$\Pi^{\mu\nu}(q)=\Pi(\qpara^2,\qperp^2)
(g_\parallel^{\mu\nu}-q^\mu_\parallel q^\nu_\parallel/\qpara^2)$.
This in turn implies that the full photon propagator takes the
following form in covariant gauges ($\xi=1$ is the Feynman gauge):
\begin{eqnarray}
\mathcal{D}^{\mu\nu}(q)&=&\frac{1}{q^2+\Pi(\qpara^2,\qperp^2)}
\left(g_\parallel^{\mu\nu}-\frac{q^\mu_\parallel
q^\nu_\parallel}{\qpara^2}\right)+\frac{g_\perp^{\mu\nu}}{q^2}
+\frac{q^\mu_\parallel q^\nu_\parallel}{q^2
\qpara^2}\nn\\
&&+(\xi-1)\frac{1}{q^2}\frac{q^\mu q^\nu}{q^2}.
\label{D}
\end{eqnarray}
The polarization function $\Pi(\qpara^2,\qperp^2)$ is given by
\begin{equation}
\Pi(\qpara^2,\qperp^2)=\frac{2\alpha}{\pi}\,
|eH|\,\exp\left(-\frac{q_\perp^2}{2|eH|}\right)
F\left(\frac{\qpara^2}{4m^2}\right),
\end{equation}
where $\alpha=e^2/4\pi$ is the fine-structure constant and
\begin{equation}
F(u)=1-\frac{1}{2u\sqrt{1+1/u}}\,\log\frac{\sqrt{1+1/u}+1}{\sqrt{1+1/u}-1}.
\end{equation}
The above result for $\Pi(\qpara^2,\qperp^2)$ agrees with those
obtained in
Refs.~\cite{Calucci:1993fi,Gusynin:1995nb,Gusynin:1998zq,Kuznetsov:2002zq,Hong:1996pv}.
The function $F(u)$ has the following asymptotic behavior:
$F(u)\simeq 0$ for $|u|\ll 1$ and $F(u)\simeq 1$ for $|u|\gg 1$.
The polarization effects modify the propagation of \emph{virtual}
photons in an external magnetic field. Whereas photons of momenta
$|\qpara^2|\ll m^2$ remain unscreened, photons of momenta
$m^2\ll|\qpara^2|\ll|eH|$ and $\qperp^2 \ll |eH|$ are screened
with a characteristic length $L=(2 \alpha |eH|/\pi)^{-1/2}$.

In terms of \eqref{G} and \eqref{Sigma}, the fermion self-energy in
the BVA, when evaluated on the fermion mass shell, is found to be
given by~\cite{Leung:2005yq}
%\begin{widetext}
\begin{eqnarray}
m\,\Delta[\sgn(eH)]&=&ie^2\intq\,\exp\left(-\frac{q_\perp^2}{2|eH|}\right)
%\nn\\&&\times
\gammapara^\mu\,\frac{1}{\gammapara\cdot(p-q)_\parallel+m}
\,\gammapara^\nu\,\nn\\
&&\times\mathcal{D}_{\mu\nu}(q)\,\Delta[\sgn(eH)]\bigg|_{\ppara^2=-m^2},\label{m}
\end{eqnarray}
where $\mathcal{D}_{\mu\nu}(q)$ is given by \eqref{D} and use has
been made of \eqref{Delta1} and \eqref{Delta2}. The presence of
$\Delta[\sgn(eH)]$ in \eqref{m} is again a consequence of the
LLLA. Using the WT identity in the BVA \eqref{WT2}, one can rewrite
the \emph{would-be} gauge dependent contribution (denoted symbolically
as $\Sigma_\xi$) on the right-hand side of \eqref{m} as
\begin{eqnarray}
\Sigma_\xi&=&ie^2(\xi-1)(\gammapara\cdot\ppara+m)\intq\,
\exp\left(-\frac{q_\perp^2}{2|eH|}\right)\frac{1}{(q^2)^2}\nn\\
&&\times \frac{1}{\gammapara\cdot(p-q)_\parallel+m}\,
\gammapara\cdot\qpara\,\Delta[\sgn(eH)].\label{mxi}
\end{eqnarray}
%\end{widetext}
Since $\Sigma_\xi$ is proportional to $(\gammapara \cdot \ppara +
m)$, it vanishes identically on the fermion mass shell
$\ppara^2=-m^2$ or, equivalently, $\gammapara\cdot\ppara+m=0$.
This, together with the transversality of the vacuum polarization,
completes our proof that the BVA is a consistent truncation of the
SD equations. Consequently, the dynamical fermion mass, obtained
as the solution of the truncated SD equations evaluated on the
fermion mass shell, is manifestly gauge independent.

Having proved the gauge independence of the dynamical fermion mass
in the BVA, we are now ready to find $m$ by solving \eqref{m}
self-consistently. Note that the transverse components in
$\mathcal{D}^{\mu\nu}(q)$ decouple in the LLLA. Following the same
argument given above in the proof of the on-shell gauge
independence, one can verify that contributions from the
longitudinal components in $\mathcal{D}^{\mu\nu}(q)$ proportional to
$q^\mu_\parallel q^\nu_\parallel/\qpara^2$ vanish identically on the
fermion mass shell. Therefore, only the first term in
$\mathcal{D}^{\mu\nu}(q)$ proportional to $g_\parallel^{\mu\nu}$
contributes to the on-shell SD equation \eqref{m}. Consequently, the
matrix structures on both sides of \eqref{m} are consistent. With
this, we find from \eqref{m} the gap equation that determines the
dynamical fermion mass~\cite{Leung:2005yq}
\begin{eqnarray}
m&=& \frac{\alpha}{2\pi^2}\int
d^2\qpara\frac{m}{q_3^2+(q_4-m)^2+m^2}
%\nn\\&&\times
\int_0^\infty d\qperp^2
\frac{\exp(-q_\perp^2/2|eH|)}{\qpara^2+\qperp^2+\Pi(\qpara^2,\qperp^2)},
\label{gap}
\end{eqnarray}
where $\qpara^2 = q_3^2 + q_4^2$. Here we have made a Wick rotation
to Euclidean space and used the mass shell condition
$\ppara^\mu=(m,0)$.

\begin{figure}[t]
\begin{center}
\includegraphics[width=4.0in,keepaspectratio=true]{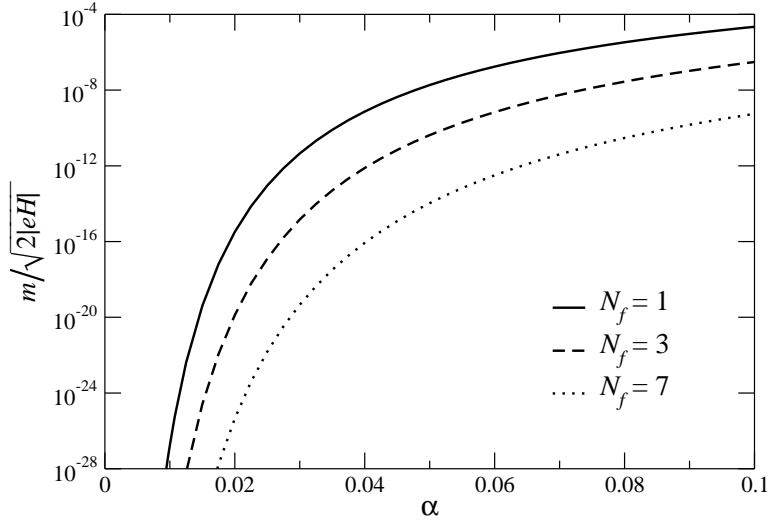}
\end{center}
\caption{Plot of $m$ as a function of $\alpha$ for several values
of $N_f$.} \label{fig}
\end{figure}

The generalization of our result to the case of QED with $N_f$
fermion flavors can be done straightforwardly by the replacement
$\Pi(\qpara^2,\qperp^2)\to N_f\,\Pi(\qpara^2,\qperp^2)$ in
\eqref{D} and \eqref{gap}. We have numerically solved \eqref{gap}
to obtain $m$ as a function of $\alpha$ for several values of
$N_f$ (see Fig.~\ref{fig}). Numerical analysis shows that the
solution of \eqref{gap} can be fit by the following analytic
expression:
\begin{equation}
m=a\,\sqrt{2|eH|}\;\beta(\alpha)\,
\exp\left[-\frac{\pi}{\alpha\log(b/N_f\,\alpha)}\right],
\end{equation}
where $a$ is a constant of order one, $b\simeq 2.3$, and $\beta(\alpha)
\simeq N_f\,\alpha$.

Our result differs from those obtained in
Refs.~\cite{Gusynin:1998zq,Kuznetsov:2002zq}, in which the so-called
improved rainbow approximation is used. We note that the improved
rainbow approximation used in
Refs.~\cite{Gusynin:1998zq,Kuznetsov:2002zq} is exactly the same as
the BVA used in this article. It can be verified fairly easily that
the truncated SD equations in both approximations resum identically
the same infinite subset of diagrams. Before concluding this
article, we argue that to a large extent those earlier results can
be attributed to gauge dependent artifacts. A detailed comparison to
previous works and further discussions will be presented
elsewhere~\cite{Leung:2005yq}.

The authors of Ref.~\cite{Gusynin:1998zq} claimed that (i) in
covariant gauges there are one-loop vertex corrections arising from
the term $\qpara^\mu\qpara^\nu/q^2\qpara^2$ in the full photon
propagator that are not suppressed by powers of $\alpha$ (up to
logarithms) and hence need to be accounted for; (ii) there exists a
noncovariant and nonlocal gauge in which, and only in which, the BVA
is a reliable truncation of the SD equations that consistently
resums these one-loop vertex corrections. The gauge independent
analysis in the BVA, as presented in this article, shows clearly
that such contributions vanish identically on the fermion mass
shell.  To put it another way, had the authors of
Ref.~\cite{Gusynin:1998zq} calculated properly the physical,
on-shell dynamical fermion mass (as we have done in our study),
they would not have found the ``large vertex corrections'' they
obtained, and therefore their claim that the BVA (or improved
rainbow approximation) is a good approximation only in the special
noncovariant and nonlocal gauge they invoke is not valid.
Together with the fact that in the BVA there are no diagrams
with vertex corrections being resummed by the SD equations, our
analysis calls into question about the validity of their
conclusions.

We emphasize that the WT identity is a necessary condition for the
gauge independence of the dynamical fermion mass, but it is far from
sufficient. While the WT identity guarantees the truncated vacuum
polarization is transverse, it guarantees only the truncated
on-shell fermion self-energy is gauge independent. Hence, the
dynamical fermion mass is gauge independent only when determined by
the position of the fermion pole obtained in a consistent
truncation. This is tantamount to evaluating the truncated fermion
self-energy on the fermion mass shell. Even though the WT identity
in the BVA is verified in Ref.~\cite{Gusynin:1998zq}
in a particular noncovariant and nonlocal gauge, this does not
guarantee that the dynamical fermion mass obtained therein from the
truncated fermion self-energy evaluated off the fermion mass shell
will be gauge independent. In fact, the particular noncovariant and
nonlocal gauge is invoked by hand such that the gauge dependent
contribution cancels contributions from terms proportional
to $\qpara^\mu \qpara^\nu/\qpara^2$ in the full photon
propagator. Our gauge independent analysis in the
BVA reveals clearly that such a gauge fixing is not only ad hoc and
unnecessary, but also leaves the issue of gauge independence
unaddressed.

In a recent article~\cite{Kuznetsov:2002zq}, the authors claimed
that (i) in QED with $N_f$ fermion flavors a critical number
$N_{cr}$ exists for any value of $\alpha$, such that chiral symmetry
remains unbroken for $N_f >N_{cr}$; (ii) the dynamical fermion mass
is generated with a double splitting for $N_f< N_{cr}$. As can be
gleaned clearly from Fig.~\ref{fig}, both these conclusions are
incorrect. They are gauge dependent artifacts of an inconsistent
truncation. On the one hand, the SD equation for the fermion
self-energy was obtained (in an unspecified ``appropriate'' gauge)
in the BVA \emph{within} the LLLA. On the other hand, the vacuum
polarization was calculated in the BVA but \emph{beyond} the LLLA.
This, however, is not a consistent truncation of the SD equations because
the WT identity in the BVA can be satisfied only within the
LLLA~\cite{Leung:2005yq}.

The result of Ref.~\cite{Kuznetsov:2002zq} suggests that in the
inconsistent truncation as well as in the unspecified gauge used
there, the unphysical, gauge dependent contributions from higher
Landau levels are so large that they become dominant over the
physical, gauge independent contribution from the LLL and lead to
the authors' incorrect conclusions. Hence, we emphasize that the LLL
dominance in a strong magnetic field should be understood in the
context of consistent truncation schemes as follows. Contributions
to the dynamical fermion mass from higher Landau levels that are
obtained in a (yet to be determined) consistent truncation of the SD
equations are subleading when compared to that from the LLL obtained
in the consistent BVA truncation. To the best of our knowledge, such
a consistent truncation has not appeared in the literature.

In conclusion, we have presented a consistent truncation of the SD
equations in the LLLA that allows us to study in a gauge independent
manner the physics of chiral symmetry breaking in a strong external
magnetic field. The gauge independent approach to the
Schwinger-Dyson equations discussed in this article is general in
nature, and hence not specific to the problem at hand. We believe
that this approach will be useful in other areas of physics that
also require a nonperturbative understanding of gauge theories.

S.-Y.W.\ is grateful to E.\ Mottola for those long discussions
they had on issues of gauge (in)dependence that a year later made
this work a reality. We would like to thank V.\ A.\ Miransky for
helpful correspondence. This work was supported in part by the US
Department of Energy under grant DE-FG02-84ER40163.

\end{document}